\documentclass[12pt,A4]{article}
\usepackage {amsmath}
\usepackage {epsfig}
\begin{document}
\textbf{\large Effect of spatial dispersion on the spectrum of inter-edge magnetoplasmons in the
two-dimensional heterogeneous system.}
\begin{flushright}
\vspace{0.2 cm}T.I.Mogilyuk \\
Kurchatov Research Center / Institute of Superconductivity and Solid
State Physics, Moscow, Russia
\end{flushright}
\section{Abstract}
\begin{small}
 The present paper is devoted to the theoretical study of the spectrum of
low-frequency electronic density oscillations (plasmon waves)
running along the boundary of two contacting two-dimensional
electronic systems in the perpendicular magnetic field. For the
first time, such waves were predicted in the Institute of
Radio-engineering and Electronics of RAS ( V.A. Volkov, S.A.
Mikhailov, 1992), studied experimentally in the papers of foreign
investigations and called inter-edge magnetoplasmons (IEMP). In a
specific case when the two-dimensional system is finite and has the
half-plane shape, the internal boundary becomes external and these
plasmons go over into the well-known edge magnetoplasmons.

The existing theory of edge and inter-edge magnetoplasmons is built with the neglect of spatial dispersion of
the conductivity. The dispersion becomes essential when the electric potential changes significantly over the
radius of the electron Larmor orbit or cyclotron radius $R_{c}$. Here we attempt to derive and analyze the
IEMP dispersion equation involving spatial dispersion of conductivity.

The approximations are made as follows. The inner boundary is
assumed to be a straight line at which the electron concentration
changes weakly but stepwise. The same refers to all components of
the tensor magnetoconductivity. The jump of the diagonal
conductivity is considered  small compared with that of the Hall
magnetoconductivity. The latter is neglected everywhere with the
exception of cutting nonphysical divergencies. We neglect the
spatial dispersion of the Hall conductivity. The IEMP frequency  is
small compared with the cyclotron frequency. Finally, we solve a
linearized system of quasi hydrodynamic equations of electron motion
in the self-consistent field describing small fluctuations of
concentration, potential, and current with the neglect of
retardation.

The main result is the following. The existence of the spatial
dispersion results in analogue of the so-called geometric resonance
in the shortwave limit.  The IEMP frequency as a function of wave
vector oscillates with the period determined by a ratio of electron
cyclotron radius to the plasmon wavelength. For long wavelengths,
the IEMP dispersion curve coincides with the known result.
\end{small}

\section{Introduction.}
Edge magnetoplasmons represent collective excitations in the two-dimensional electron system, which propagate
along the edge  boundary of the system and are localized beside it. The edge magnetoplasmons are
experimentally discovered on the surface of liquid helium in the magnetic field normal to the helium surface.
As is stated in $\cite{state5}$,  edge magnetoplasmons have some important properties, provoking recent
interest.

(1) They have  a gapless spectrum $\omega_{emp}\sim q\ln 1/|q|$,
$\vec{q}$ being directed along the boundary of the sample. Their
frequency is much smaller than the cyclotron one. Plasmons propagate
along the boundary in the direction depending on the angle between
the outer normal to the edge and the direction of magnetic field.
The EMP frequency  is proportional to the electron concentration and
inversely proportional to the magnetic field $B$ and the size of the
sample. In the finite sample the wavevector  is discrete $q=2\pi /P$
where $P$ is the perimeter of the sample. Depending on these
parameters, the EMP frequency  can vary within large range from
infrared frequencies in submillimetric samples of semiconductor
heterostructures, e.g., quantum dots, rings, threads, antidots, down
to microwave and even audio-frequencies (kHz) in the rarefied
two-dimensional electronic system of macroscopic scale (cm) on the
liquid helium surface.

(2) In strong magnetic fields $\omega_{c}\tau >> 1$ the EMP damping
 is very small both for $\omega_{emp}\tau >> 1$ and for $\omega_{emp}\tau <<
 1$. This is the property that gives an opportunity to observe the EMP
 experimentally at the frequencies lower than 1 GHz.

 (3) The EMP frequency is determined by the Hall motion of electrons and is proportional to the
 Hall conductivity of the two-dimensional electron
system. Since the quantum Hall effect is observed not only for
direct current but also for microwave frequencies, the EMP can be
used as a powerful instrument of studying both integer and
fractional Hall effect. So, the EMP spectroscopy can be used as a
method of studying edge electronic states that play a significant
role in creating the Hall effect.

 (4) The EMP charge  is strongly localized beside the boundary of a
sample over the length comparable with the width of edge electronic states.

 The first experimental observation of the main IEMP properties is
made in $\cite{state}$ in which the properties of electrons on the
liquid helium surface are studied. It is found that the direction of
the IEMP motion depends on the sign of the  difference between the
electron concentrations on the right-hand  and left-hand sides of
the boundary. In the limit of strong fields the frequency is
proportional to the difference in the electron concentrations and
inversely proportional to the magnitude of magnetic field. The
results are in a qualitative agreement with the theoretical
predictions.

 In $\cite{state2}$ it is found that the width of the IEMP curves increases for the temperatures below the
 melting temperature of electron Coulomb crystal.

The IEMP damping and width are determined by the mean diagonal
conductivity in the boundary region. Under crystallization the
conductivity $\sigma_{xx}$ increases drastically for moderate
intensive fields. The IEMP plasmons are localized smaller than the
EMP ones. Thus the IEMP plasmons are more  sensitive to
crystallization in the high electron density region than the EMP
ones. Such vanishing cannot be achieved for the EMP plasmons.

In work $\cite{condmat}$ it is shown that, unlike edge
magnetoplasmons, inter-edge magnetoplasmons  even in the
collisionless limit have the damping connected with emitting  bulk
plasmons to the region of lower concentration. The damping is weak
for the significant difference in the concentrations on the left and
right sides. There are two modes, upper and lower, in the IEMP
spectrum.
 The upper mode has the frequency larger than the cyclotron one.

 The upper branch has a strong nondissipative damping connected
 with the emission of bulk 2D magnetoplasmons to the region of lower
 concentration. This effect takes place only in the two-dimensional
 system.

 The damping is weak only for a significant difference in the electron
 concentrations on the right and  left sides from the boundary $N^{r}>>N^{l}$
  and large frequency
  $$\omega_{r}(q_{y})>>|\omega_{c}|\,[\omega_{r}(q_{y})=\sqrt{\frac{2\pi{N^{r}e^{2}}q_{y}}{m^{*}}}]$$
as  compared with the cyclotron one.

The lower mode does not exist for such fields at which
$\alpha\omega_{c}\tau\leq{1}$. It is well-defined for
$$\omega_{c}\tau>>\frac{1}{\alpha\;}>1,$$ where $\tau$ is the decay
time of momentum and $ \alpha=(N^{r}-N^{l})/(N^{r}+N^{l})$.

The IEMP frequency in any geometry with the unhomogeneous
conductivity is proportional to the inhomogeneity of the Hall
conductivity. The Hall conductivity is inversely proportional to the
magnitude of magnetic field. Thus the IEMP frequency  proves to be
small compared with the cyclotron one.

For calculating spectrum in the geometry with the nonuniform conductivity which changes abruptly at the
boundary of a disc, we use the method in which the diagonal conductivity is replaced by some effective
quantity  $\sigma_{eff}$ and the inhomogeneity in $\;\delta\sigma_{xx}$ is considered as a small correction.
The quantity  $\delta\sigma_{xx}$ should not be small in comparison with $\sigma_{xx}$ but only with
$\delta\sigma_{xy}$.

 The approximate method $\delta\sigma_{xx}=0$  is based on the fact that
 inhomogeneity in the diagonal conductivity has a relatively weak effect
 on the spectrum. Taking inhomogeneity into account  is necessary to
 avoid a logarithmic divergence in the dispersion equation. The averaged
 quantity $\sigma_{eff}$ determines the scale of spatial dispersion in the
 permittivity and is determined from the condition of the current
 continuity at the boundary of half-planes.

With the help of analyzing an exact solution for the geometry of two half-planes we show an applicability
 $\delta\sigma_{xx}=0$ not only for weak inhomogeneity but also it is stated that the error of finding
the spectrum at $\alpha\leq1$ does not exceed 8.5\%$\;$.

The spectrum in the geometry of two half-planes with the different conductivities reads:
\begin{equation}
\omega(q_{y})=\frac{2\delta\sigma_{xy}(\omega)q_{y}F[q_{y}{\overline{l}(\omega)}]}{\kappa},
\end{equation}
where $F(z)$ is determined in the following way:
\begin{equation}
F(z)=\int_{0}^{\frac{\pi}{2}}\frac{dt}{\sin{t}+z}=\left\{%
\begin{array}{ll}
    \ln{\frac{2}{z}},\; 0<z\ll 1, \\
    \frac{\pi}{2z},\; z\gg 1. \\
\end{array}
\right.
\end{equation}
Here
 $$\overline{l}(\omega)=\frac{l^{r}(\omega)+l^{l}(\omega)}{2},\,l^{r}(\omega)=\frac{2\pi{i\sigma_{xx}^{r}(\omega)}}
{\omega},\,l^{l}(\omega)=\frac{2\pi{i\sigma_{xx}^{l}(\omega)}}{\omega}$$ and $\kappa$ is the permittivity of
the ambient medium. It is shown that the spectrum in the system of more complicated geometry such as a square
or rectangle can be obtained with the help of quantization rule,  replacing $q_{y}$  with $2n/P$.

\section{A set of the problem}

$\;\;\,$ Let the 2D system of two half-planes with the different tensors of two-dimensional conductivity
$\sigma^{l}_{\alpha\beta}(q,\omega)$ and
 $\sigma^{r}_{\alpha\beta}(q,\omega)$ for the right and left half-planes, respectively,
is placed in a strong magnetic field.   We seek for the dispersion equation for spectrum $\omega(q_{y})$,
 $q_{y}$ being the wavevector of oscillations along the boundary of half-planes. In addition,
$\delta\sigma_{xx}(q,\omega)=|\sigma^{l}_{xx}(q,\omega)-\sigma^{r}_{xx}(q,\omega)|$,
$\,\delta\sigma_{xy}=|\sigma^{l}_{xy}-\sigma^{r}_{xy}|$. We make the following approximations:
\\
(1)  conductivity and concentration change abruptly at the boundary of half-planes;
\\
(2) $\sigma_{xy}(q,\omega)=\sigma_{xy}(0,\omega)$;
\\
(3) $\delta\sigma_{xy}=|\sigma^{l}_{xy}-\sigma^{r}_{xy}|>>\delta\sigma_{xx}$;
\\
(4) lack of retardation;
\\
(5) quasi hydrodynamic approximation.
\par
For simplicity, the effective electron mass is the same everywhere. The permittivity of the environment equals
unity. Let us introduce the coordinate axes in which the $y$-axis is directed along the boundary of
half-planes and the region $x>0$ refers to the right half-plane.
\par
We have the following equation in such coordinates:
\begin{equation}
1=\frac{q_y\delta\sigma_{yx}(0,\omega)}{\omega}\int_{-\infty}^{+\infty}{\frac{dq_{x}}{q\epsilon(q,\omega)}},
\end{equation}
where $\epsilon(q,\omega)=1+ql_{eff}(q,\omega)$,
$l_{eff}(q,\omega)=\frac{2\pi{i}\sigma_{eff}(q,\omega)}{\omega}$ ($\sigma_{eff}(q,\omega)$ will be defined
later),
 $\sigma_{xy}=-\sigma_{yx},$
$\,\,\, q=\sqrt{q_{x}^{2}+q_{y}^{2}}$.
\par
 Let us introduce the
following notations, such as: $\rho$ is the 2D density of electrons, $\vec{q}=(q_{x},q_{y})$ is the 2D
wavevector, $\vec{r}$=(x,y), $\varphi(\vec{r},t)$ is a fluctuation of the potential in  point $\vec{r}$ at
moment $t$, $\vec{j}(\vec{r},t)$ is the 2D vector of the current density, $\omega$ is the  frequency of
oscillations, $\sigma(\vec{r},{R},t)$ is the kernel of the 2D operator of conductivity,
$l^{l}(q,\omega)=2\pi{i}\sigma^{l}(q,\omega)/\omega,\;l^{r}(q,\omega)= 2\pi{i}\sigma^{r}(q,\omega)/\omega$,
$\delta{l(q,\omega)}=|l^{r}(q,\omega)-l^{l}(q,\omega)|,
\;\delta{N}=|N^{r}-N^{l}|,\;\delta{J_{0}^{2}(qR)}=|J_{0}^{2}(qR^{r})-J_{0}^{2}(qR^{l})|$, $m^{*}$ is the
effective electron mass, $a^{*}_{B}$ is the Bohr radius of an electron with effective mass $m^{*}$, and
$R^{l},R^{r}$ is the cyclotron radius of the left and right half-planes, respectively.

\section{Derivation of the dispersion equation.}

\par We use the following equations: \\ (1)
continuity equation of current in the Fourier representation:
\begin{equation}
\vec{q}\vec{j}(\vec{q},\omega)=\omega{\rho(\vec{q},\omega)},
\end{equation}
(2) the 2D Poisson formula in the Fourier representation:
\begin{equation}
\varphi(\vec{q},\omega)=\frac{2\pi{\rho(\vec{q},\omega)}}{q}
\end{equation}
(3) the 2D Ohm law:
\begin{equation}
\vec{j}(\vec{r},t)=-\int_{-\infty}^{+\infty}\int_{-\infty}^{+\infty}\sigma_{\alpha\beta}(\vec{r},\vec{R},t)\frac{\partial\varphi(\vec{R},t)}{\partial{x_{\beta}}}d^{2}\vec{R},
\end{equation}
Neglecting both the spatial dispersion of the Hall conductivity and the inhomogeneity of diagonal
conductivity, we obtain:
\begin{equation}
\left\{%
\begin{array}{ll}
    j_{x}(\vec{r},\omega)=-\int_{-\infty}^{+\infty}\int_{-\infty}^{+\infty}\sigma_{eff}(\vec{r},\vec{R},\omega)\frac{\partial\varphi(\vec{R},\omega)}{\partial{x}}d^{2}\vec{R}-\sigma_{xy}(\vec{r},\omega)\frac{\partial\varphi(\vec{r},\omega)}{\partial{y}}, \\
    j_{y}(\vec{r},\omega)=-\int_{-\infty}^{+\infty}\int_{-\infty}^{+\infty}\sigma_{eff}(\vec{r},\vec{R},\omega)\frac{\partial\varphi(\vec{R},\omega)}{\partial{y}}d^{2}\vec{R}+\sigma_{xy}(\vec{r},\omega)\frac{\partial\varphi(\vec{r},\omega)}{\partial{x}}.  \\
\end{array}%
\right.
\end{equation}
\begin{equation}
\left\{%
\begin{array}{ll}
    j_{x}(\vec{r},\omega)=-\int_{-\infty}^{+\infty}\int_{-\infty}^{+\infty}\sigma_{eff}(\vec{r},\vec{R},\omega)\frac{\partial\varphi(\vec{R},\omega)}{\partial{x}}d^{2}\vec{R}-\sigma_{xy}(\vec{r},\omega)\frac{\partial\varphi(\vec{r},\omega)}{\partial{y}}, \\
    j_{y}(\vec{r},\omega)=-\int_{-\infty}^{+\infty}\int_{-\infty}^{+\infty}\sigma_{eff}(\vec{r},\vec{R},\omega)\frac{\partial\varphi(\vec{R},\omega)}{\partial{y}}d^{2}\vec{R}+\sigma_{xy}(\vec{r},\omega)\frac{\partial\varphi(\vec{r},\omega)}{\partial{x}}.  \\
\end{array}%
\right.
\end{equation}
Assuming that
$\sigma_{xy}(\vec{r},\omega)=-\sigma_{yx}(\vec{r},\omega)=\sigma^{r}_{xy}(\omega)\theta(x)+\sigma^{l}_{xy}(\omega)\theta(-x)$,
one may find the current in the Fourier components :
\begin{equation}
\left\{%
\begin{array}{ll}
j_{x}(\vec{q},\omega)=-i\sigma_{eff}(\vec{q},\omega)q_{x}\varphi(\vec{q},\omega)-\frac{iq_{y}}{\sqrt{2\pi}}\int_{-\infty}^{+\infty}{\exp{(iq_{x}x)}}\sigma_{xy}(\omega)\varphi(x,q_{y})\,
dx \: ,\\
j_{y}(\vec{q},\omega)=-i\sigma_{eff}(\vec{q},\omega)q_{y}\varphi(\vec{q},\omega)+\frac{iq_{x}}{\sqrt{2\pi}}\int_{-\infty}^{+\infty}{\exp{(iq_{x}x)}\sigma_{xy}(\omega)\varphi(x,q_{y})\,
dx}-\\-\frac{\delta\sigma_{xy}(\omega)\varphi(\vec{q},\omega)}{\sqrt{2\pi}}\;
.
\end{array}%
\right.
\end{equation}
Substituting this into the continuity equation, one arrives at:
\begin{equation}
\delta\sigma_{xy}(0,\omega)q_{y}\varphi(x=0,q_{y},\omega)=\frac{q\omega\varphi(\vec{q},\omega)}{\sqrt{2\pi}}[1+\frac{2\pi{i}\sigma_{xx}(q,\omega)q}{\omega}].
\end{equation}
Performing the inverse Fourier transform in $q_{x}$ at $x=0$, we have:
\begin{equation}
\omega(q_{y})=q_{y}\delta\sigma_{yx}(0,\omega)\int_{-\infty}^{+\infty}{\frac{dq_{x}}{q\epsilon_{eff}(q,\omega)}}.
\end{equation}

\section{Analysis of the dispersion equation in the limiting cases and the diagram of the IEMP spectrum.}

\par
We take the following expression for Hall conductivity $\sigma_{yx}$ from $\cite{state1}$. For $qR\ll 1$,
\begin{equation}
\sigma_{yx}(q,\omega)=\frac{Ne^{2}}{m^{*}\omega_{c}}\left[\frac{1}{1-\frac{\omega^{2}}{\omega_{c}^{2}}}+
\Sigma_{n=2}^{\infty}\left(\frac{qR}{2}\right)^{2n-2}\frac{n(\frac{\omega^{2}}{\omega_{c}^{2}})}{(n!)^{2}(n^{2}-
\frac{\omega^{2}}{\omega_{c}^{2}})}].\right.
\end{equation}
It is evident that $\sigma_{yx}(q,0)=\frac{Ne^{2}}{m^{*}\omega_{c}}$. Then we evaluate the magnitude of the
region for wave vectors q in which we can neglect the  spatial dispersion. Let $\omega /\omega_{c}=0.5$ and in
the case $qR\sim{13}$ the second term of a Fourier series for the Hall conductivity depends  on q. We use the
known expression:
\begin{equation}
\epsilon(q,\omega)=1+\frac{4{m^{*}e^{2}}}{q\hbar^{2}}\Sigma_{m=1}^{\infty}\frac{m^{2}J_{m}^{2}(qR)}{m^{2}-\frac{\omega^{2}}{\omega_{c}^{2}}},
\end{equation}
where $R$ is some effective cyclotron radius of the system. For $\omega /\omega_{c}\ll 1$, this expression
significantly simplifies:
\begin{equation}
\epsilon(q,0)=1+\frac{2}{qa^{*}_{B}}[1-J_{0}^{2}(qR)],
\end{equation}
with  the help of
\begin{equation}
1=J_{0}^{2}(qR)+2\Sigma_{m=1}^{\infty}J_{m}^{2}(qR).
\end{equation}
For $qR\gg 1$, $\epsilon(q,\omega)\sim1$ and thus the integral in the dispersion equation is logarithmic. In
order to avoid divergence for small $q_{y}\ll 1/R$, we neglect the spatial dispersion of diagonal
conductivity:
\begin{equation}
\epsilon_{eff}(q,\omega)=1+ql_{eff}(0,\omega),
\end{equation}
\begin{equation}
l_{eff}(0,0)=\frac{l^{r}+l^{l}}{2}=\frac{R^{2}}{a^{*}_{B}},
\,R^{2}=\frac{R^{2\,r}+R^{2\,l}}{2}.
\end{equation}
For large $q_{y}\gg 1/R$, we cut the upper bound at $\frac{1}{\delta{l(q_{y},0)}}$. Then we have the following
spectrum $\omega(q_{y})$:
\begin{equation}
\frac{\omega(q_{y})}{\omega_{c}}=\left\{%
\begin{array}{ll}
        \frac{4\delta{N}q_{y}{e^{2}}}{m^{*}\omega_{c}^{2}\sqrt{1-(\frac{q_{y}R^{2}}{a^{*}_{B}})^{2}}}\rm{arctanh}{\sqrt{\frac{(a^{*}_{B}-q_{y}R^{2})}{(a^{*}_{B}+q_{y}R^{2})}}},\;\textrm{при}\;\frac{q_{y}R^{2}}{a^{*}_{B}}<1, \\
        \frac{4\delta{N}q_{y}{e^{2}}}{m^{*}\omega_{c}^{2}\sqrt{(\frac{q_{y}R^{2}}{a^{*}_{B}})^{2}-1}}\arctan{\sqrt{\frac{(q_{y}R^{2}-a^{*}_{B})}{(q_{y}R^{2}+a^{*}_{B})}}}, \;\textrm{при}\;\frac{q_{y}R^{2}}{a^{*}_{B}}>1,  \\
        \frac{2q_{y}\delta{N}e^{2}}{m^{*}\omega_{c}^{2}}\int_{0}^{\frac{q_{y}^{2}a^{*}_{B}}{2\delta{J^{2}_{0}(q_{y}R)}}}{\frac{dq_{x}}{q+\frac{2}{a^{*}_{B}}}},\;\textrm{при}\;q_{y}R>>1. \\
\end{array}%
\right.
\end{equation}
For $\frac{1}{l_{eff}(0,0)}\ll q_{y}\ll\frac{1}{R}$, $\frac{\omega(q_{y})}{\omega_{c}}$ asymptotically tends
to $\frac{\delta{N}}{2N_{eff}}$. When $q_{y}\gg\frac{1}{R}$,  peaks are observed due to vanishing
$\delta{l_{eff}(q_{y})}=0$ at some $q_{y}$.
\par
To plot the dispersion curve, we take the following values in the CGS system for the constants used :
$m^{*}=0.2*9.1*10^{-28}\textrm{g},\; a^{*}_{B}=25*10^{-9}\textrm{cm},\;N^{l}=2*10^{12}\textrm{cm}^{2},\;
N^{r}=2.4*10^{12}\textrm{cm}^{2},\;\omega_{c}=2*10^{11}\textrm{s}^{-1}.$

\section{Conclusion}

\par
We have studied the effect of spatial dispersion on the spectrum of inter-edge magnetoplasmons. From the Fig.2
one can see that the periodic oscillations are observed for $q_{y}\sim n/R$. For $q_{y}\ll 1/R$, the spectrum
coincides with that in the lack of the spatial dispersion of diagonal conductivity. The physical reason for
the resonances in the IEMP spectrum  is the cyclotron resonance when the frequency of electron rotation in the
magnetic field equals the IEMP frequency or differs by a multiple factor from it.
\begin{figure}
\includegraphics[width=16cm,height=16cm,angle=0]{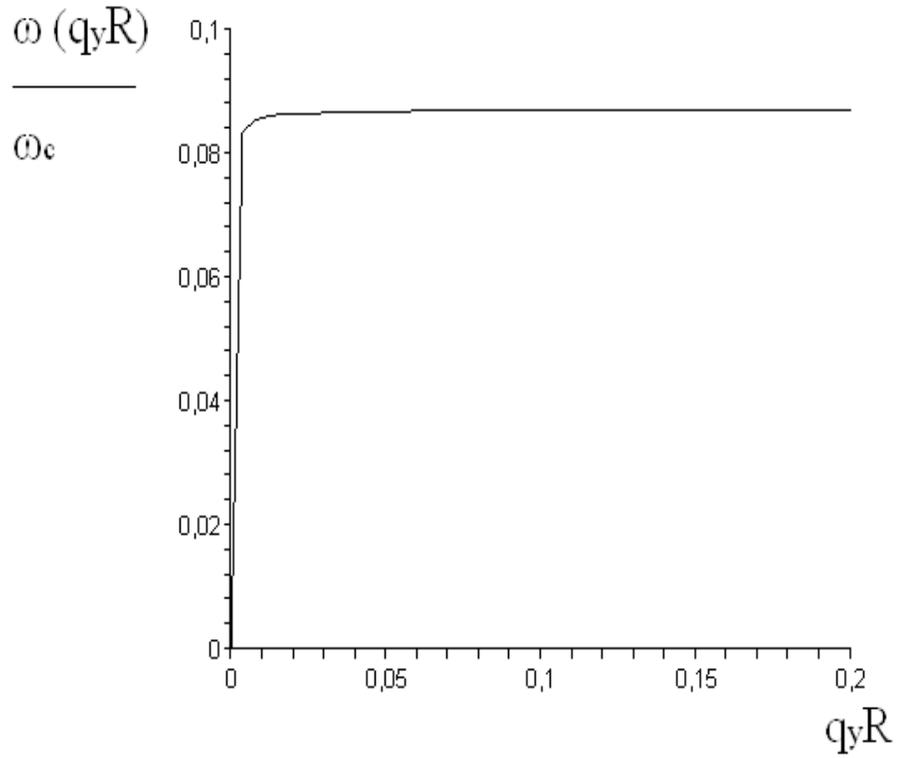}
\caption{The IEMP dispersion curve $\omega(q_{y}R)/\omega_{c}$ with involving spatial dispersion in the system
of two half-planes with inhomogeneous conductivity $q_{y}R\ll 1$. The $y$-axis represents dimensionless
quantity $\omega(q_{y}R)/\omega_{c}$.}
\end{figure}
\begin{figure}
\includegraphics[width=18cm,height=15cm,angle=0]{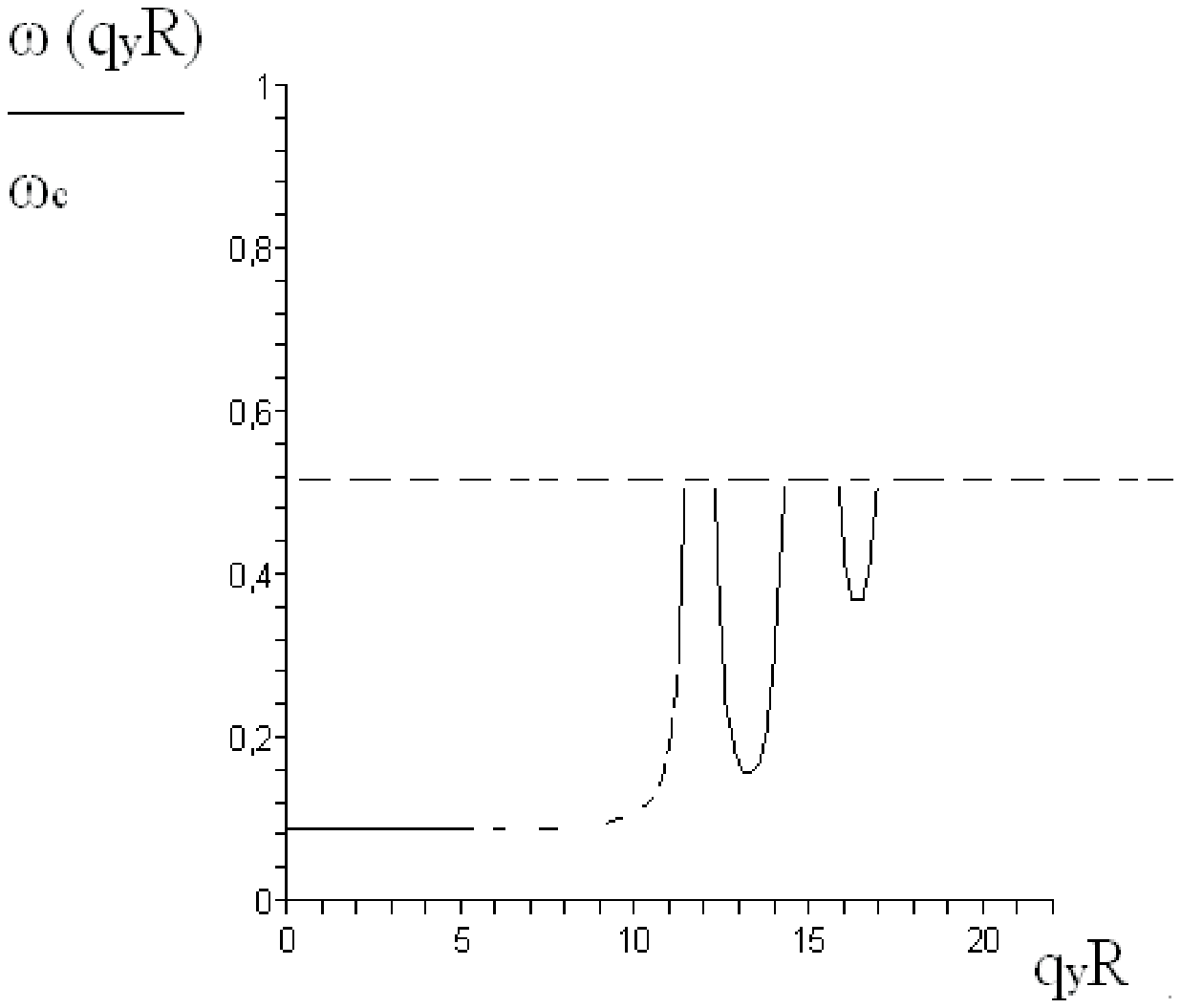}
\caption{The IEMP dispersion curve $\omega(q_{y}R)/\omega_{c}$  with involving spatial dispersion in the
system of two half-planes with the unhomogeneous conductivity. The $y$-axis represents dimensionless quantity
$\omega(q_{y}R)/\omega_{c}$.}
\end{figure}

\newpage
\end{document}